\documentclass[onecolumn,showpacs,preprintnumbers,elsart]{revtex4}
\usepackage{amssymb}
\usepackage{amsmath}
\usepackage{mathrsfs}
\usepackage{graphicx}
\usepackage{dcolumn}
\usepackage{bm}
\usepackage{color}
\usepackage[center]{subfigure}

\begin{document}

\title{Two-dimensional solitons in dipolar Bose-Einstein condensates with
spin-orbit coupling}
\author{Xunda Jiang$^{1}$}
\author{Zhiwei Fan$^{1}$}
\author{Zhaopin Chen$^{1,2}$}
\author{Wei Pang$^{3}$}
\author{Yongyao Li$^{1}$}
\email{yongyaoli@gmail.com}
\author{Boris A. Malomed$^{2}$}
\affiliation{$^{1}$Department of Applied Physics, College of Electronic Engineering,
South China Agricultural University, Guangzhou 510642, China \\
$^{2}$ Department of Physical Electronics, School of Electrical Engineering,
Faculty of Engineering, Tel Aviv University, Tel Aviv 69978, Israel.\\
$^{3}$Department of Experiment Teaching, Guangdong University of Technology,
Guangzhou 510006, China}

\begin{abstract}
We report families of two-dimensional (2D) composite solitons in spinor
dipolar Bose-Einstein condensates, with two localized components linearly
mixed by the spin-orbit coupling (SOC), and the intrinsic nonlinearity
represented by the dipole-dipole interaction (DDI) between atomic magnetic
moments polarized in-plane by an external magnetic field. Recently, stable
solitons were predicted in the form of \textit{semi-vortices} (composites
built of coupled fundamental and vortical components) in the 2D system
combining the SOC and contact attractive interactions. Replacing the latter
by the anisotropic long-range DDI, we demonstrate that, for a fixed norm of
the soliton, the system supports a \emph{continuous family} of stable
spatially asymmetric vortex solitons (AVSs), parameterized by an offset of
the pivot of the vortical component relative to its fundamental counterpart.
The offset is limited by a certain maximum value, while the energy of the
AVS practically does not depend on the offset. At small values of the norm,
the vortex solitons are subject to a weak oscillatory instability. In the
present system, with the Galilean invariance broken by the SOC, the
composite solitons are set in motion by a kick whose strength exceeds a
certain depinning value. The kicked solitons feature a negative effective
mass, drifting along a spiral trajectory opposite to the direction of the
kick. A critical angular velocity, up to which the semi-vortices may follow
rotation of the polarizing magnetic field, is found too.
\end{abstract}

\pacs{03.75.Lm; 05.45.Yv; 03.75.Mn}
\maketitle






\section{Introduction}

The creation of stable two- and three-dimensional (2D and 3D) self-trapped
modes (bright solitons and solitary vortices) in various physical media,
including optical waveguides and atomic Bose-Einstein condensates (BECs),
had been identified as a challenging problem some time ago \cite{review,12},
retaining this status up to now. A fundamental difficulty is that, while the
ubiquitous cubic self-focusing nonlinearity, which represents the Kerr
effect in optics \cite{Agrawal} and attractive inter-atomic collisions in
BEC [the latter may be controlled by means of the Feshbach resonance (FR)
\cite{Randy,Bauer2009,Yan2013}], readily creates stable solitons in the 1D
geometry, all localized multidimensional modes supported by the cubic
self-focusing are unstable. They are destroyed by the critical \cite{7,8,op1}
or supercritical \cite{8,op1} collapse, which is driven by the cubic
attractive nonlinearity in the 2D and 3D, respectively. 2D and 3D bright
solitons with embedded vorticity are subject to a still stronger instability
against splitting by perturbations breaking their axial symmetry \cite%
{review,12}.

A possibility for the stabilization of multidimensional solitons is
suggested by using self-focusing nonlinearity weaker than cubic. Indeed,
quadratic (alias second-harmonic-generating) nonlinearity does not cause
collapse in 2D and 3D geometries, hence it may be used for making stable
fundamental solitons, as demonstrated experimentally in the 2D spatial
domain \cite{Torruellas}, and shown, both theoretically \cite{20} and
experimentally \cite{21}, for 2D spatiotemporal \textquotedblleft optical
bullets" \cite{21}. However, the quadratic nonlinearity does not remove the
azimuthal splitting instability of vortex solitons \cite{22,op2}.

Both fundamental and vortex bright solitons may be made stable using
combinations of competing self-focusing and self-defocusing nonlinear terms,
namely, cubic-quintic \cite{23-25,op3,op4,op5} or quadratic-cubic \cite{26}.
In these settings, all the fundamental solitons are stable (indeed, the
creation of 2D stable spatial solitons in a colloidal optical waveguide,
featuring the cubic-quintic nonlinearity, has been recently reported \cite%
{Falcao}), while vortex solitons are stabilized above a specific threshold,
which actually implies that they must be very broad, making their
experimental creation difficult.

The most straightforward means for the stabilization of fundamental and
vortical solitons in 2D and 3D geometries is provided by the use of
spatially periodic potentials. In optical media, such potentials can be
imposed in the form of virtual photonic lattices in photorefractive crystals
\cite{Moti}, and permanent lattices written in bulk silica \cite{Szameit},
while in BEC similar potentials are induced by interference patterns
(optical lattices) created by laser beams illuminating the condensate \cite%
{28}. In the experiment, photonic-lattice potentials were used to make
stable 2D fundamental and vortex optical solitons \cite{29,op6,op7,op8}, as
well as 3D spatiotemporal \textquotedblleft bullets" \cite{61}. A recent
experimental result is the creation of 2D plasmon-polariton solitons in
microcavities, which are also supported by a lattice structure \cite{30}.

Another possibility for the stabilization of 2D fundamental solitons is
offered by \textquotedblleft management" techniques, i.e., periodic
alternation of the sign of the nonlinearity \cite{35}. This possibility was
originally proposed for the transmission of optical beams across a medium
built as a periodic alternation of layers of self-focusing and defocusing
materials \cite{32}. The relevance of the same method for the stabilization
of 2D matter-wave solitons in BEC was demonstrated too, with the help of
periodic switch of the sign of the cubic nonlinearity by means of the FR
\cite{33,op9,op10}. However, the management cannot stabilize 2D solitary
vortices, nor 3D fundamental solitons.

A completely different approach to the creation of self-trapped fundamental
and vortex modes was recently proposed in Ref. \cite%
{37,op11,op12,others,op13}. It relies on the use of\emph{\ self-defocusing}
nonlinearity, whose local strength in the space of dimension $D$ grows from
the center to periphery, as a function of distance $r$, at any rate faster
than $r^{D}$, supporting extremely robust families of solitons, solitary
vortices, and more complex modes, such as \textit{hopfions}, which carry two
different topological charges \cite{Yasha}, for $D=1,2,3$.

A new possibility was recently revealed by the theoretical analysis of
two-component (spinor) BEC with the linear spin-orbit coupling (SOC) between
the components, that may be induced by a combination of appropriate optical
and magnetic fields \cite{SOC}-\cite{Evgeny}. In most cases, the SOC is
considered in the combination with the repulsive collisional nonlinearity,
which may give rise to delocalized (\textquotedblleft dark") vortices in the
2D setting \cite{SOC-vortex,op16,op17,op18,op19} and gap solitons of the
bright type, if a lattice potential is included \cite{SOC-gap,op20,op21}. It
is also possible to consider the interplay of the SOC with the intrinsic
attractive nonlinearity. The latter setting easily supports 1D solitons \cite%
{1D-SOC-solitons,op22,op23}. An unexpected result was recently produced by
the consideration of the 2D self-attractive SOC system: while it was
previously believed that any system with the cubic self-focusing would give
rise to unstable solitons in the free 2D space, in view of the simultaneous
occurrence of the \textit{critical collapse} \cite{8,op1,7}, the SOC system
(in particular, of the Rashba type) creates \emph{stable} solitons of
\textit{semi-vortex} and \textit{mixed-mode} types, with one fundamental ($%
\psi _{+}$) and one vortex ($\psi _{-}$) wave-function components, or a
mixture of fundamental and vortex modes in both components, respectively
\cite{Sakaguchi,Sakaguchi2,Wesley}. Although the SOC terms in the respective
Gross-Pitaevskii equations (GPEs) break the rotational invariance [see Eq. (%
\ref{basiceq}) below], the two components of the semi-vortex solution, with
chemical potential $\mu $, feature the axial symmetry in polar coordinates $%
\left( r,\theta \right) $,%
\begin{equation}
\psi _{+}\left( r,\theta ,t\right) =e^{-i\mu t}f_{1}\left( r^{2}\right)
,~\psi _{-}\left( r,\theta ,t\right) =e^{-i\mu t+i\theta }rf_{2}\left(
r^{2}\right) ,  \label{semi}
\end{equation}%
while the mixed modes do not maintain this symmetry, and cannot be
represented by an exact ansatz similar to Eq. (\ref{semi}) \cite{Sakaguchi}.
This ansatz, as well as more complex mixed-mode solutions (they are stable
if the attraction between $\psi _{+}$ and $\psi _{-}$ is stronger than the
self-attraction of the components), clearly demonstrate that the SOC makes
two components of the soliton's spinor wave function topologically
different. Mobile 2D mixed-mode solitons were found too, which is a
nontrivial issue, as the SOC breaks the Galilean invariance of the system
\cite{Sakaguchi}. Furthermore, the same stabilization mechanism was
elaborated for 2D spatiotemporal solitons in a dual-core planar optical
waveguide, with SOC emulated by dispersive linear coupling between the two
cores \cite{Yaro}. An explanation to these findings is provided by the fact
that the dimensionality of the SOC coefficient defines a spatial scale in
the system; thus, it breaks the scaling invariance of the underlying 2D\
GPEs with the cubic terms, and lifts the related degeneracy of the norm of
the two-component solitons, placing them \emph{below} the threshold for the
onset of the collapse. Being thus protected against the collapse, the
self-trapped 2D solitons become stable objects, which realize the ground
state of the 2D setting.

It is easier to predict stable multidimensional solitons in media with
nonlocal cubic nonlinearities, where the collapse does not occur. In optics,
the nonlocality is featured by reorientational nonlinearity of liquid
crystals \cite{Peccianti,op24,op25}, as well as by thermal response of a
dielectric medium \cite{Wieslaw,op26,op27}. In BEC, effective nonlocal
nonlinearities originate from the isotropic Van der Waals interactions
between Rydberg atoms \cite{Heidemann}, and long-range dipole-dipole
interactions (DDIs) of atoms carrying permanent magnetic dipole moments \cite%
{Griesmaier,Mlu,Aikawa,dipolar-review}. In the latter context, stable 1D
bright and dark solitons \cite{Cuevas,Bland2015}, 2D bright solitons \cite%
{Pedir}, 3D dark solitons \cite{Nath}, and bright solitary vortices \cite%
{Tikhonenkov} were predicted, assuming that attractive DDI can be
(artificially) realized in isotropic configurations. On the other hand,
fundamental and vortex solitons, in the 1D and 2D geometry alike, may be
readily made stable in the natural setting with repulsive DDIs between
field-induced (rather than permanent) atomic moments oriented perpendicular
to the system's plane, if the strength of the polarizing field (magnetic or
electric) grows from the center to periphery faster than $r^{3}$ \cite{we}.
In a more realistic anisotropic 2D setting with the \emph{in-plane}
polarization of permanent atomic moments, stable fundamental
(zero-vorticity) solitons were predicted too \cite{Tikhonenkov2,op28}. A
challenging problem, which was not explored before, is a possibility of
existence of \textit{anisotropic solitary vortices}, i.e., anisotropic
bright solitons, supported by the DDI, with an embedded topological charge
that may be interpreted as vorticity.

The objective of the present work is to construct stable 2D topological
solitons in the model of the spinor BEC with the in-plane polarization of
atomic moments, combining the linear SOC (of the Rashba type) and nonlinear
long-range DDI. The SOC-DDI system was introduced, in a form relevant to the
experimental implementation, in Refs. \cite{SOC-DDI,op29,op30}. 2D
fundamental solitons in the system of this type, which includes the
attractive DDI and local repulsive nonlinearity, were recently investigated
in Ref. \cite{Xuyong} (to focus on the most interesting situation, we here
consider only the DDI nonlinearity). Modes of two kinds were found in the
latter work, \textit{viz}., smooth solitons with a spatially varying phase,
and stripe solitons with a spatially oscillating density, similar to those
which were reported in Ref. \cite{Sakaguchi2}, that was dealing with a
combination of the SOC, attractive local nonlinearity, and lattice
potentials in 2D. Mobility of the 2D solitons was also addressed in Ref.
\cite{Xuyong}.

In this work, we report stable 2D topological solitons, in the form of
spatially symmetric and asymmetric semi-vortices. The asymmetry is
represented by separation between the pivot of the vortex component and
central point of its fundamental counterpart, which breaks the system's
reflectional symmetry,
\begin{equation}
(x,y)\rightarrow (-x,-y),~\left( \psi _{+},\psi _{-}\right) \rightarrow
\left( \psi _{+},-\psi _{-}\right) .  \label{inversion}
\end{equation}
To the best of our knowledge, this is the first example of stable asymmetric
vortex solitons (AVSs) found in any nonlinear system. In particular, they
are completely different from azimuthons found in isotropic media with local
\cite{azi,op31,op32,op33} or nonlocal \cite{azi-nonloc} interactions.
Simultaneously, the AVSs offer the first example of 2D vortical solitons
supported by the anisotropic in-plane DDI.

The paper is structured as follows. The model is introduced in Section II.
Basic results for stationary symmetric vortex solitons (SVSs) and their
mobility are presented in Section III. In Section IV, we consider AVSs,
which, as said above, feature a symmetry-breaking shift between the vortex
and fundamental components. The possibility of setting SVSs and AVSs into
rotation is addressed in Section V. The paper is concluded by section VI.

\section{The model}

In the mean-field approximation, the binary dipolar BEC is governed by the
system of coupled GPEs for the spinor wave function, $\psi =(\psi _{+},\psi
_{-})$, written here in the scaled 2D form:
\begin{gather}
i{\frac{\partial \psi _{+}}{\partial t}}=-{\frac{1}{2}}\nabla ^{2}\psi
_{+}+\lambda \left( {\frac{\partial \psi _{-}}{\partial x}}-i{\frac{\partial
\psi _{-}}{\partial y}}\right)  \notag \\
+\psi _{+}(\mathbf{r})\int \int d\mathbf{r^{\prime }}R(\mathbf{r}-\mathbf{%
r^{\prime }})\left[ |\psi _{+}(\mathbf{r^{\prime }})|^{2}+|\psi _{-}(\mathbf{%
r^{\prime }})|^{2}\right] ,  \notag \\
i{\frac{\partial \psi _{-}}{\partial t}}=-{\frac{1}{2}}\nabla ^{2}\psi
_{-}-\lambda \left( {\frac{\partial \psi _{+}}{\partial x}}+i{\frac{\partial
\psi _{+}}{\partial y}}\right)  \notag \\
+\psi _{-}(\mathbf{r})\int \int d\mathbf{r^{\prime }}R(\mathbf{r}-\mathbf{%
r^{\prime }})\left[ |\psi _{-}(\mathbf{r^{\prime }})|^{2}+|\psi _{+}(\mathbf{%
r^{\prime }})|^{2}\right] .  \label{basiceq}
\end{gather}%
Here $\lambda $ is the Rashba-SOC coefficient, the contact nonlinearity is
neglected, as said above, and the DDI kernel is
\begin{equation}
R(\mathbf{r}-\mathbf{r^{\prime }})={\frac{1-3\cos ^{2}\Theta }{[\epsilon
^{2}+(\mathbf{r}-\mathbf{r^{\prime }})^{2}]^{3/2}}},  \label{DDI}
\end{equation}%
where $\epsilon $ is a regularization scale, which is determined by the
transverse size of the nearly-2D layer \cite{Cuevas,HJS}. The form of this
kernel implies that the dipoles are polarized, by an external magnetic
field, in the 2D plane along the positive direction of the $x$ axis, hence $%
\cos ^{2}\Theta \equiv \left( x-x^{\prime }\right) ^{2}/\left\vert \mathbf{r}%
-\mathbf{r}^{\prime }\right\vert ^{2}$. Adequate results, with a
characteristic size of 2D solitons being much larger than $\epsilon $
(otherwise, the solitons are not quasi-2D objects), can be produced, e.g.,
with $\epsilon =0.05$, if $\lambda =1$ is set by scaling. These values of $%
\epsilon $ and $\lambda $ are fixed below.

An essential peculiarity of Eqs. (\ref{basiceq}) and (\ref{DDI}) is the
interplay of two different kinds of the spatial anisotropy, which are
introduced, severally, by the linear SOC and nonlinear DDI terms. While the
DDI anisotropy is characterized by the form of kernel (\ref{DDI}), the
structure of the SOC Hamiltonian of the Rashba type, acting on spinor $%
\left( \psi _{+}.\psi _{-}\right) ^{\mathrm{T}}$ in Eq. (\ref{basiceq}), is
elucidated by writing it in the polar coordinates:%
\begin{equation}
\mathbf{H}_{\mathrm{SOC}}^{\mathrm{(Ra)}}=\lambda \left( \sigma _{x}\hat{p}%
_{y}-\sigma _{y}\hat{p}_{x}\right) =\lambda \left(
\begin{array}{cc}
0 & e^{-i\theta }\left( \frac{\partial }{\partial r}-\frac{i}{r}\frac{%
\partial }{\partial \theta }\right) \\
-e^{+i\theta }\left( \frac{\partial }{\partial r}+\frac{i}{r}\frac{\partial
}{\partial \theta }\right) & 0%
\end{array}%
\right) ,  \label{H}
\end{equation}%
where $\mathbf{p}=-i\nabla $ is the momentum operator, and $\sigma _{x,y}$
are the Pauli matrices. In particular, this structure makes it clear why
ansatz (\ref{semi}) produces the exact solution in the model with the
contact interactions.

The remaining symmetries of Eq. (\ref{basiceq}) correspond to the invariance
with respect to the specular reflections in the $x$ and $y$ directions:%
\begin{eqnarray}
x &\rightarrow &-x,~t\rightarrow -t,~\psi _{\pm }\rightarrow \pm \psi _{\pm
}^{\ast },  \label{xx} \\
y &\rightarrow &-y,~t\rightarrow -t,~\psi _{\pm }\rightarrow \psi _{\pm
}^{\ast },  \label{yy}
\end{eqnarray}%
where $\ast $ stands for the complex conjugate, and, independently, to the
invariance with respect to the flip of the two components of the (pseudo-)
spinor wave function:%
\begin{equation}
\psi _{+}\rightleftarrows \psi _{-},~x\rightarrow -x.  \label{flip}
\end{equation}%
In particular, the above-mentioned inversion transformation, defined as per
Eq. (\ref{inversion}), is a product of reflections (\ref{xx}) and (\ref{yy}).

Stationary solitons are introduced as usual,
\begin{equation}
\left\{ \psi _{+}(\mathbf{r},t),\psi _{-}(\mathbf{r},t)\right\} =\left\{
\phi _{+}(\mathbf{r}),\phi _{-}(\mathbf{r})\right\} e^{-i\mu t},
\label{solitonsolution}
\end{equation}%
with total norm
\begin{equation}
N=N_{+}+N_{-}=\int \int |\psi _{+}(\mathbf{r})|^{2}d\mathbf{r}+\int \int
|\psi _{-}(\mathbf{r})|^{2}d\mathbf{r},  \label{N}
\end{equation}%
and energy
\begin{equation}
E=E_{\mathrm{kin}}+E_{\mathrm{SOC}}+E_{\mathrm{DDI}},  \label{Energy}
\end{equation}%
\begin{gather}
E_{\mathrm{kin}}={\frac{1}{2}}\int \int d\mathbf{r}(|\nabla \phi
_{+}|^{2}+|\nabla \phi _{-}|^{2}),  \notag \\
E_{\mathrm{SOC}}={\frac{\lambda }{2}}\int \int d\mathbf{r}\left[ \phi
_{+}^{\ast }\left( {\frac{\partial \phi _{-}}{\partial x}}-i{\frac{\partial
\phi _{-}}{\partial y}}\right) +\phi _{-}^{\ast }\left( -{\frac{\partial
\phi _{+}}{\partial x}}-i{\frac{\partial \phi _{+}}{\partial y}}\right) +%
\mathrm{c.c}\right] ,  \notag \\
E_{\mathrm{DDI}}={\frac{1}{2}}\int \int d\mathbf{r}d\mathbf{r^{\prime }}R(%
\mathbf{r}-\mathbf{r^{\prime }})\left[ |\phi _{+}(\mathbf{r})|^{2}+|\phi
_{-}(\mathbf{r})|^{2})(|\phi _{+}(\mathbf{r^{\prime }})|^{2}+|\phi _{-}(%
\mathbf{r^{\prime }})|^{2}\right] .  \label{Eddi}
\end{gather}%
$N$ and $E$ are dynamical invariants of the system, along with the linear
momentum, $i\int \int d\mathbf{r}(\phi _{+}\nabla \phi _{+}^{\ast }+\phi
_{-}\nabla \phi _{-}^{\ast })$ [it remains conserved, in spite of the
breaking of the Galilean invariance by the SOC terms in Eq. (\ref{basiceq}%
)], while the conservation of the angular momentum is broken by both the SOC
and DDI terms.

\section{Symmetric vortex solitons (SVSs)}

\subsection{Quiescent solitons}

Unlike the SOC model with contact attractive interactions, the anisotropic
DDI\ in Eq. (\ref{basiceq}) does not admit any exact ansatz for
semi-vortices, cf. Eq. (\ref{semi}). A variational approximation for the
stationary states can be developed, but it turns out to be very cumbersome.
Therefore, stationary SVS solutions were found by means of the
imaginary-time-integration method \cite{Chiofalo,Jianke}, and their
stability was tested by means of real-time simulations. All the numerical
results were produced using\textbf{\ }square-shaped numerical grids with
mesh sizes $\delta x=\delta y$. The input for the imaginary-time simulations
was taken as a semi-vortex, using the above-mentioned ansatz (\ref{semi}) as
a pattern:
\begin{equation}
\phi _{+}^{(0)}=A_{+}\exp (-\alpha _{+}r^{2}),\quad \phi
_{-}^{(0)}=A_{-}re^{i\theta }\exp \left( -\alpha _{-}r^{2}\right) ,
\label{guess1}
\end{equation}%
where $A_{1,2}$ and $\alpha _{1,2}>0$ are real constants\textbf{.} In the
course of the evolution in imaginary time, the mode was deformed, but its
topological structure persisted.

\begin{figure}[tbp]
\centering{\label{fig1a} \includegraphics[scale=0.35]{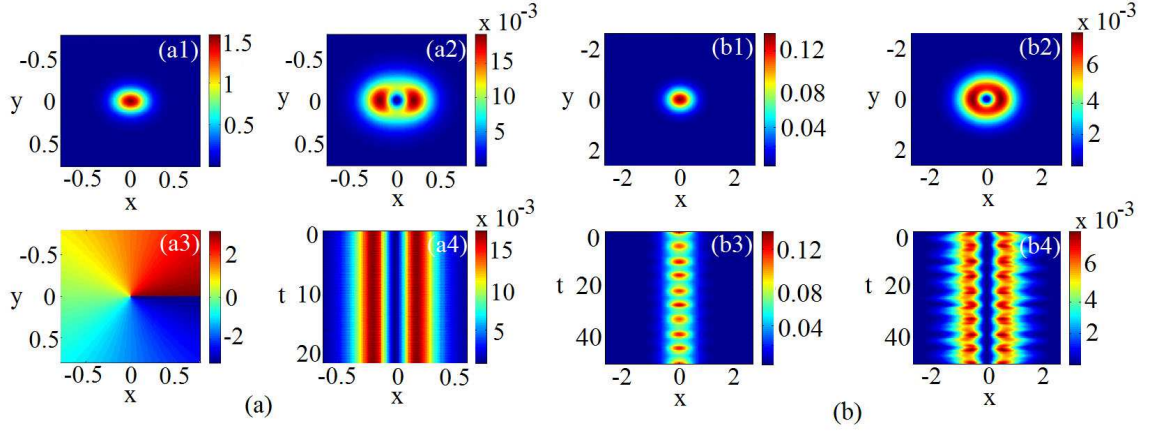}}
\caption{(Color online) (a) The stable SVS with total norm $N=0.15$. (a1,a2)
Intensity distributions of the fundamental and vortical components, $|%
\protect\phi _{+}(x,y)|^{2}$ and $|\protect\psi _{-}(x,y)|^{2}$,
respectively. (a3) The phase structure of the vortical component. (a4)
Direct real-time simulations corroborating the stability of this SVS (the
evolution of the vortical component is displayed in cross section $y=0$).
The evolution time in this panel exceeds $100$ characteristic diffraction
times of the soliton. (b) An unstable SVS with $N=0.10$. (b1,b2) The same as
in (a1,a2), for the unstable SVS. (b3,b4) Real-time simulations show the
oscillatory instability of both components, in cross section $y=0$, by means
of the density contour plots. The resulting breathers keeps the vorticity
and symmetries of the original SVS. }
\label{ExpSVS}
\end{figure}

Numerically generated results demonstrates that the solitons are stable if
the total norm exceeds a certain threshold, $N>N_{\mathrm{th}}\approx 0.12$.
A typical example of the output in the form of \emph{stable} SVSs with norm $%
N=0.15$ is displayed in Fig. \ref{ExpSVS}(a). In this figure, it is clearly
seen that both components feature a typical anisotropic structure, being
stretched along the horizontal axis, which is easily explained by the fact
that DDI\ kernel (\ref{DDI}) implies the attractive interaction along $x$,
i.e., for $\Theta =0$ and, more generally, for $\cos ^{2}\Theta <1/3$ \cite%
{Tikhonenkov2,op28}. On the other hand, the density patterns in both
components of the semi-vortex are symmetric with respect to the $x$ and $y$
axes [i.e., they are invariant with respect to transformations (\ref{xx})
and (\ref{yy})], that is why this soliton is called SVS. When the norm below
$N_{\mathrm{th}}$, the semi-vortex develops an oscillatory instability,
which converts it into a breather. Nevertheless, the original topological
structure and symmetries of the SVS are held by the robust breather, as
shown in Fig. \ref{ExpSVS}(b3,b4); in addition to the oscillations displayed
in that figure, the elliptically deformed vortex component ($\psi _{-}$)\
performs periodic \textquotedblleft nutations", with its axes oscillating
around the average directions.

It is relevant to stress that such a stability threshold for the semi-vortex
solitons, in terms of the norm, is different from the stability conditions
found in the system combining the SOC and contact attractive interactions,
where the solitons remain stable from arbitrarily small values of $N$ up to
the collapse threshold \cite{Sakaguchi}. In that system, the stability is
determined by the fact that the spatial scale, introduced by the SOC, breaks
the scaling invariance of the 2D GPE, pushing the soliton's norm below the
collapse-onset threshold, thus securing the stability against the collapse
\cite{Sakaguchi}. For the system with the nonlocal interaction considered
here, the stability is, in a sense, easier to achieve, as the long-range DDI
by itself does not lead to the collapse.

\begin{figure}[tbp]
\centering{\label{fig2a} \includegraphics[scale=0.2]{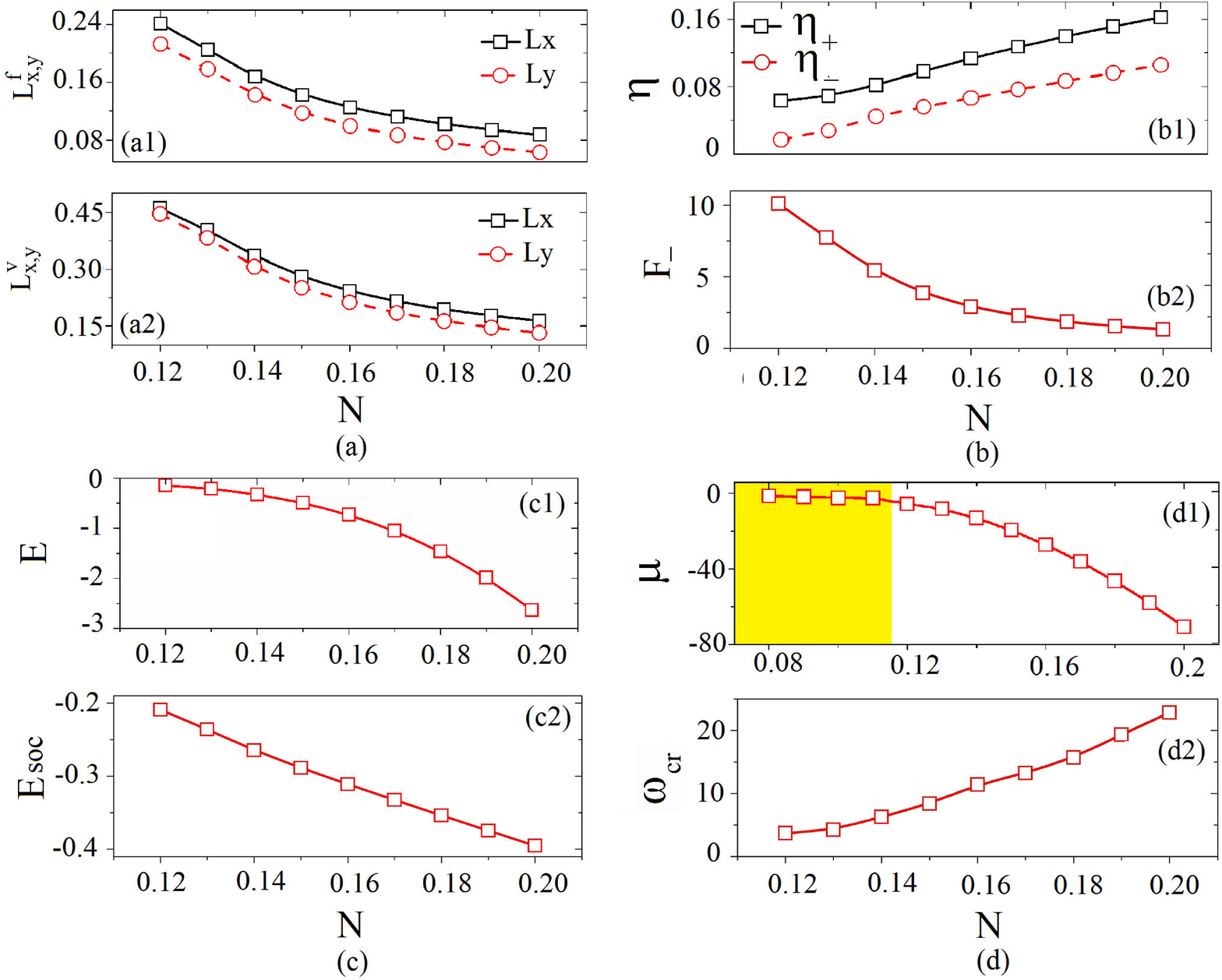}}
\caption{(Color online) (a1,a2) Horizontal and vertical sizes [defined per
Eq. (\protect\ref{Aeff})] versus $N$ for the two SVS\ components. (b1)
Anisotropies of both components, $\protect\eta _{\pm }$, versus $N$ [defined
as per Eq. (\protect\ref{eta})]. (b2) The norm share of the vortical
component of the SVS, $F_{-}$, versus $N$ [defined as per Eq. (\protect\ref%
{F-factor})]. (c1,c2) The total energy of the SVS, and the energy of the
spin-orbit coupling, $E_{\mathrm{SOC}}$, versus $N$. (d1) The chemical
potential of the SVS, $\protect\mu $, versus $N$. In the yellow area,
stationary SVSs are subject to the oscillatory instability, transforming
into breathers which keep the same structure as the SVSs. (d2) The critical
rotation speed, $\protect\omega _{\mathrm{cr}}$, versus $N$. }
\label{ProSVS}
\end{figure}

To quantify properties of the entire SVS family, we define the following
characteristics of these self-trapped modes:

\noindent (1) Effective horizontal ($\mathrm{effx}$) and vertical ($\mathrm{%
effy}$) sizes of the fundamental, \textquotedblleft f" ($\phi _{+}$), and
vortex, \textquotedblleft v" ($\phi _{-}$), components: 

\begin{equation}
{\left\{ L_{\mathrm{effx}}^{\mathrm{f,v}},L_{\mathrm{effy}}^{\mathrm{f,v}%
}\right\} =\left[ {\frac{{\int \int \left\{ \left( x-\mathrm{X}_{\mathrm{mc}%
}^{\mathrm{f,v}}\right) ^{2},\left( y-\mathrm{Y}_{\mathrm{mc}}^{\mathrm{f,v}%
}\right) ^{2}\right\} |\phi _{+,-}(x,y)|^{2}dxdy}}{\int \int |\phi
_{+,-}(x,y)|^{2}dxdy}}\right] ^{1/2},}  \label{Aeff}
\end{equation}%
where $\left\{ \mathrm{X}_{\mathrm{mc}}^{\mathrm{f,v}},\mathrm{Y}_{\mathrm{mc%
}}^{\mathrm{f,v}}\right\} $ are coordinates of the center of mass (c.o.m.)
for the two components, which are defined as
\begin{equation}
{\left\{ \mathrm{X}_{\mathrm{mc}}^{\mathrm{f,v}},\mathrm{Y}_{\mathrm{mc}}^{%
\mathrm{f,v}}\right\} ={\frac{{\int \int \left\{ x,y\right\} |\phi
_{+,-}(x,y)|^{2}dxdy}}{{\int \int |\phi _{+,-}(x,y)|^{2}dxdy}}}}.
\label{com}
\end{equation}

\noindent (2) The anisotropy of each component:
\begin{equation}
\eta _{+,-}={\frac{\left\vert L_{\mathrm{effx}}^{\left( \mathrm{f,v}\right)
}-L_{\mathrm{effy}}^{\left( \mathrm{f,v}\right) }\right\vert }{{L_{\mathrm{%
effx}}^{\left( \mathrm{f,v}\right) }+L_{\mathrm{effy}}^{\left( \mathrm{f,v}%
\right) }}}}.  \label{eta}
\end{equation}

\noindent (3) Norm shares of the fundamental and vortex components:
\begin{equation}
F_{+,-}=\left( N_{+,-}/N\right) \times 100\%.  \label{F-factor}
\end{equation}

These characteristics, as well as the chemical potential [see Eq. (\ref%
{solitonsolution})] and the total energy [see Eq. (\ref{Energy})] of the SVS
are displayed versus the total norm in Fig. \ref{ProSVS}, at $N\geq 0.12$,
where, as said above, the stationary SVSs are stable. Figure \ref{ProSVS}(a)
demonstrates that, quite naturally, the SVS shrinks with the increase of $N$%
, its size in the $x$-direction being larger than along $y$, and the
anisotropy degree, $\eta _{+,-}$, increases with $N$, see Fig. \ref{ProSVS}%
(b1). This finding can be easily explained by the fact that the increase of $%
N$ leads to the domination of the nonlinear DDI, which tends to stretch 2D
solitons along the $x$ axis, as mentioned above. Further, Fig. \ref{ProSVS}%
(b2) demonstrates that the norm share of the vortex component, $F_{-}$,
decreases with the growth of $N$, similar to what was found for stable
semi-vortices in the system with the contact attraction \cite{Sakaguchi}. In
Fig. \ref{ProSVS}(c), the total energy of the SVS increases with $N$,
whereas the SOC energy term \ decreases, hence the DDI energy plays the
dominant role at large $N$. Finally, in Fig. \ref{ProSVS}(d1), the chemical
potential decreases with $N$, hence the SVS family satisfies the
Vakhitov-Kolokolov (VK) criterion, $d\mu /dN<0$, which is a well-known
necessary stability condition for any soliton family supported by attractive
interactions \cite{VK,8,op1}. The weak oscillatory instability of the SVS in
the yellow (shaded) area of Fig. \ref{ProSVS}(d1) is not detected by the VK
condition, which is sensitive only to instability accounted for by
non-oscillatory perturbation modes.

\subsection{Mobility of the symmetric vortex solitons}

Mobility of the SVS is a nontrivial issue, as the SOC terms break the
Galilean invariance of the underlying equations (\ref{basiceq}). In this
subsection, the mobilities of the soliton are studied by applying a kick to
a stationary soliton solution,
\begin{equation}
\psi _{\pm }(\mathbf{r},t=0)=\phi _{\pm }(\mathbf{r})\exp (i\mathbf{k}\cdot
\mathbf{r}),  \label{kick}
\end{equation}%
where $\mathbf{k}=k_{x}\mathbf{i}+k_{y}\mathbf{j}$ is the vector of the
kick. To study the resulting motion of the soliton, we define the trajectory
of the moving soliton by time-dependent coordinates of the c.o.m of the
fundamental component, as
\begin{equation}
\left\{ {\mathrm{X}_{\mathrm{mc}}^{\mathrm{f}}(t),\mathrm{Y}_{\mathrm{mc}}^{%
\mathrm{f}}(t)}\right\} {={\frac{{\int \int \left\{ x,y\right\} |\psi
_{+}(x,y,t)|^{2}dxdy}}{{\int \int |\psi _{+}(x,y,t)|^{2}dxdy}}}.}
\label{com2}
\end{equation}%
Actually, the location of the c.o.m. is dominated by the fundamental
component of the SVS, as the vortex components carries a much smaller norm,
see Fig. \ref{ProSVS}(b2). The mobility of the kicked solitons were studied
through the shape of the c.o.m. trajectories, as produced by real-time
simulations of Eq. (\ref{basiceq}).

\begin{figure}[tbp]
\centering{\includegraphics[scale=0.3]{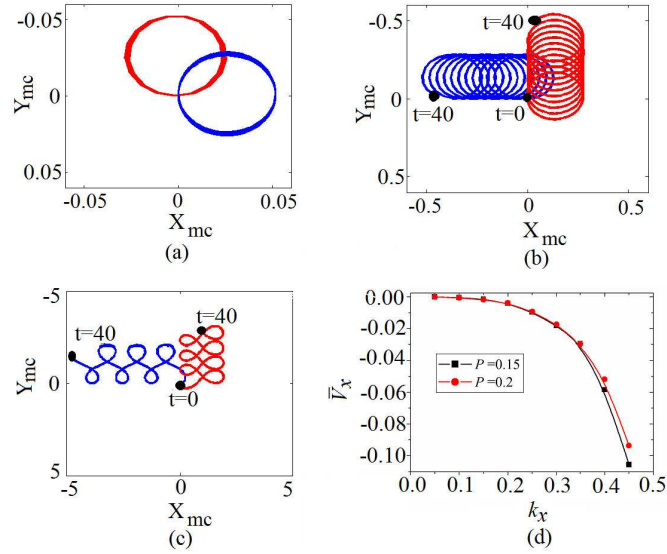}}
\caption{(Color online) (a,b,c) Trajectories of the fundamental c.o.m. of
the soliton kicked by $\mathbf{k}\equiv (k_{x},k_{y})=(0.05,0)$ (blue cycle)
\& $(0,0.05)$ (red cycle) [panel (a)], $(0.25,0)$ (blue spiral) \& $(0,0.25)$
(red spiral) [panel (b)] and $(0.5,0)$ (blue lacy curve) \& $(0,0.5)$ (red
lacy curve) [panel (c)], respectively. In all the cases, the soliton's norm
is $N=0.15$. In these panels, the evolution time is $t=40$. Panel (d)
displays the mean velocity of the spiral motion of the soliton versus $k_{x}$
for $N=0.15$ and $N=0.2$, respectively. $\bar{V}_{x}<0$ means that the
soliton moves against the direction of the applied kick.}
\label{Trajectory}
\end{figure}

Three types of the motion have been identified, for different strengths of
the kick applied along the $x-$ and $y-$directions. Typical examples of them
are displayed in Fig. \ref{Trajectory}(a,b,c). When the strength of the kick
is smaller than a \textit{depinning threshold}, the resulting trajectory is
a circle, see a typical example in Fig. \ref{Trajectory}(a), which may be
realized as produced by an effective Lorentz force acting upon the vortex
component of the soliton. Up to the depinning threshold, the radius of the
circular trajectory remains much smaller than the proper size of the
soliton. The magnitude of the threshold is $|\mathbf{k}|\approx 0.07$, which
is nearly constant in the range of $0.12<N<0.2$.

At $|\mathbf{k}|>0.07$, the depinning of the soliton occurs, transforming
the trajectory into a spiral, see a typical example in Fig. \ref{Trajectory}%
(b). In this case, the direction of the systematic drift of the soliton is
opposite to the kick, hence this type of the dynamics may be considered as a
superposition of progressive motion with a negative effective mass and
circular motion. At $|\mathbf{k}|$ exceeding a higher threshold value, the
spiral trajectory is replaced by a complex one, as shown in Fig. \ref%
{Trajectory}(c). The magnitude of the upper threshold is $|\mathbf{k}|=0.46$%
, which is also nearly constant for $0.12<N<0.2$.

At $|\mathbf{k}|$ keeps growing, the increase of the absolute value of the
drift velocity gives rise to an additional threshold value of the kick, at
which the superposition of the drift with the circular motion changes the
shape of spiral trajectory into a \textquotedblleft lacy" one, as shown in
Fig. \ref{Trajectory}(c). The magnitude of this threshold is $|\mathbf{k}%
|=0.46$, which is also nearly constant for $0.12<N<0.2$. In Fig. \ref%
{Trajectory}(d), the mean velocity of the spiral motion of the solitons with
$N=0.15$ and $0.2$ are shown vs. the size of the kick, the dependences being
very close for both values of $N$.

In Fig. \ref{Trajectory}(a,b), shifts of the soliton along the $x$ and $y$
directions, observed at $t=40$, are nearly equal, implying that its mobility
is nearly isotropic. However, in Fig. \ref{Trajectory}(c) it is seen that,
for the kick with $|k|=0.5$, which is larger than $0.46$, the $x$ and $y$
shifts, $\Delta x\approx 4.8$ and $\Delta y\approx 2.8$, are different,
which means that the mobility is anisotropic for the strong kick.

Although it may seem that the motion of the solitons observed in Fig. \ref%
{Trajectory}(b,c) violate the momentum conservation, the total vectorial
(2D) momentum is actually conserved, in all the cases. To verify the
momentum conservation, we define its normalized value as
\begin{equation}
{\mathbf{M(t)}={\frac{i\int \int \left( \psi _{+}(t)\nabla \psi _{+}^{\ast
}(t)+\psi _{-}(t)\nabla \psi _{-}^{\ast }(t)\right) dxdy}{\int \int \left(
|\psi _{+}(t)|^{2}+|\psi _{-}(t)|^{2}\right) dxdy}}.} \label{Mom}
\end{equation}%
The numerical results demonstrate that the so computed momentum remains
constant, keeping the initial value which is equal to the kick applied to
the soliton, $M=k$. An explanation to the complex motion of the soliton in
Fig. \ref{Trajectory}(b,c) is the presence of small-amplitude radiation
field, which plays the role of recoil which helps to conserve the total
momentum.\textbf{\ }

\section{Asymmetric vortex solitons (AVSs)}

\subsection{Quiescent solitons }

\begin{figure}[tbp]
\centering{\label{fig6a} \includegraphics[scale=0.25]{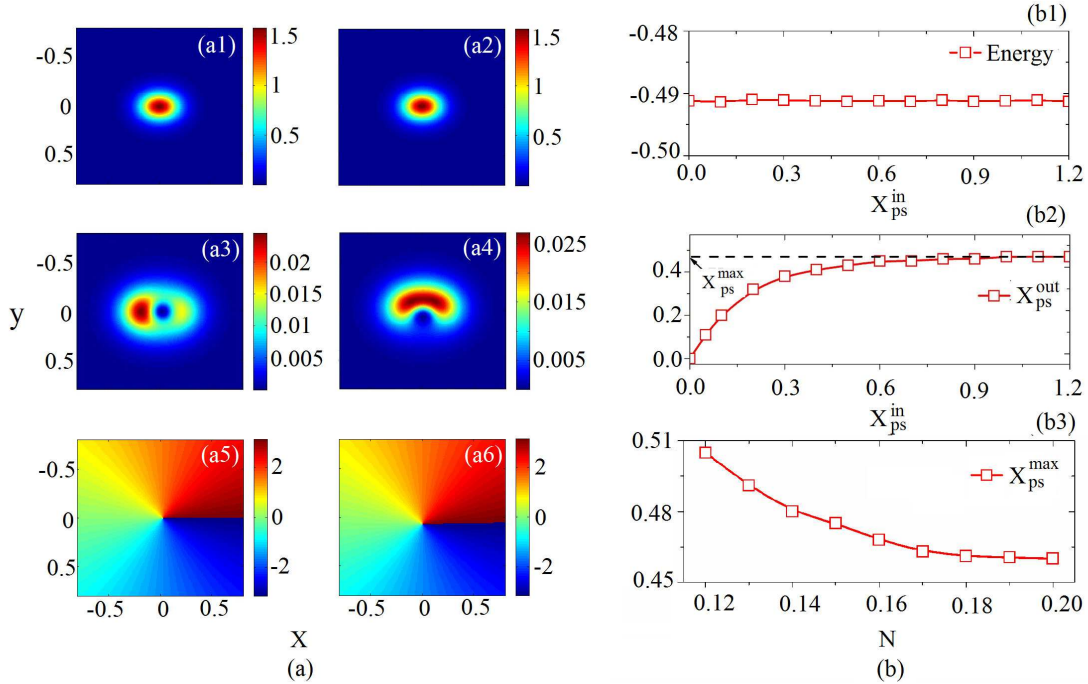}}
\caption{(Color online) (a1-a4) Typical examples of the density profile in
the fundamental (a1,a2) and vortical (a3,a4) component, $\protect\phi _{+}$
and $\protect\phi _{-}$, of a stable AVS, for the initial offset of the
vortex' pivot with respect to the fundamental component $(X_{\mathrm{ps}}^{%
\mathbf{in}},Y_{\mathrm{ps}}^{\mathbf{in}})=(0.01,0)$ and $(0,0.01)$,
respectively. The origin of the coordinate system is placed at the final
position of the pivot. The total norm of the asymmetric semi-vortex soliton
is $N=0.15$. (a5,a6) The phase structure of the vortical components ($%
\protect\psi _{-}$) from panels (a3) and (a4), respectively. The output
values of the pivot's offset of these two AVSs are $(X_{\mathrm{ps}}^{%
\mathrm{out}},Y_{\mathrm{ps}}^{\mathrm{out}})=(0.04,0)$ and $(0,0.065)$,
respectively. (b1) The energy of the AVS versus $X_{\mathrm{ps}}^{\mathrm{in}%
}$, showing the degeneracy of the family. (b2) $X_{\mathrm{ps}}^{\mathrm{out}%
}$ versus $X_{\mathrm{ps}}^{\mathrm{in}}$, the saturation taking place at
the largest output offset $X_{\mathrm{ps}}^{\mathrm{max}}=0.47$, for the AVS
with $N=0.15$. (b3) The largest output offset of the vortex' pivot along the
$x$-axis, $X_{\mathrm{ps}}^{\mathrm{out}}$, versus $N$.}
\label{ASVSs}
\end{figure}

The model with the DDI\ interactions offers a completely novel feature,
which was not found for semi-vortices in the 2D\ SOC system with the contact
attractive interactions \cite{Sakaguchi}: a possibility to produce stable
AVSs, with the asymmetry represented by spatial separation of the
fundamental and vortical components. To produce such solutions in the
quiescent form, we solved Eq. (\ref{basiceq}) by means of the
imaginary-time-evolution method, with input
\begin{equation}
\phi _{+}^{(0)}=A_{1}\exp (-\alpha _{1}r^{2}),\quad \phi _{-}^{(0)}=A_{2}|%
\mathbf{r}-\mathbf{R_{\mathrm{ps}}}|\exp (i\theta _{\mathbf{r_{\mathrm{ps}}}%
}-\alpha _{2}r^{2}),  \label{guessASVSs}
\end{equation}%
cf. Eq. (\ref{guess1}), where $\mathbf{R_{\mathrm{ps}}}=X_{\mathrm{ps}}^{%
\mathrm{in}}\mathbf{\hat{\imath}}+Y_{\mathrm{ps}}^{\mathrm{in}}\mathbf{\hat{%
\jmath}}$ is the initial offset of the vortex components with respect to the
fundamental one, and $\theta _{\mathbf{r_{\mathrm{ps}}}}$ is the angular
coordinate for the vortex' pivot placed at $\mathbf{r}=\mathbf{R_{\mathrm{ps}%
}}$ (\textquotedblleft $\mathrm{ps}$" means \textquotedblleft pivot shift").
Typical examples of the so generated stable AVSs for $N=0.15$ are displayed
in Fig. \ref{ASVSs}(a). In this figure, when we chose $(X_{\mathrm{ps}}^{%
\mathrm{in}},Y_{\mathrm{ps}}^{\mathrm{in}})=(0.01,0)$ and $(0,0.01)$, the
eventual (output) position of the pivot of the vortical component is at $(X_{%
\mathrm{ps}}^{\mathrm{out}},Y_{\mathrm{ps}}^{\mathrm{out}})=(0.04,0)$ and $%
(0,0.065)$, respectively. To produce the relationship between $(X_{\mathrm{ps%
}}^{\mathrm{out}},Y_{\mathrm{ps}}^{\mathrm{out}})$ and $(X_{\mathrm{ps}}^{%
\mathrm{in}},Y_{\mathrm{ps}}^{\mathrm{in}})$, we plot $X_{\mathrm{ps}}^{%
\mathrm{out}}$ versus $X_{\mathrm{ps}}^{\mathrm{in}}$ in Fig. \ref{ASVSs}%
(b2) (the respective dependence for the offset along $y$ is not shown here,
as it it is almost identical to that produced by the offset in the $x$
direction). The figure shows that $X_{\mathrm{ps}}^{\mathrm{out}}$ saturates
to $0.47$ with the increase of $X_{\mathrm{ps}}^{\mathrm{in}}$, in the case
of $N=0.15$. Further, the maximum output offset along $x$-axis, $X_{\mathrm{%
ps}}^{\mathrm{max}}$, i.e., the saturation value of the offset, is displayed
as a function of $N$ in Fig. \ref{ASVSs}(b3). It decreases with the increase
of $N$ because the semi-vortex itself shrinks with the growth of $N$. At $%
N<0.12$, the AVSs takes the form of a robust breather, similar to what is
shown above for the SVS.

\begin{figure}[tbp]
\centering{\label{fig8a} \includegraphics[scale=0.35]{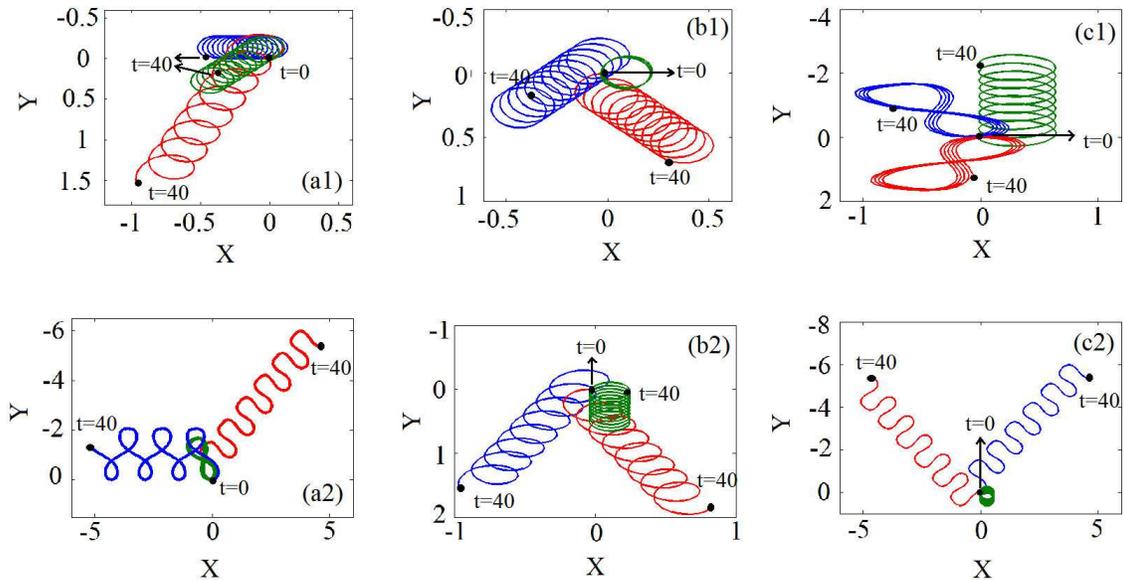}}
\caption{(Color online) (a1,a2) Trajectories of AVSs, with $\left( X_{%
\mathrm{ps}}^{\mathrm{out}},Y_{\mathrm{ps}}^{\mathrm{out}}\right) =(0,0)$
(blue curves), $(0.12,0)$ (green curves) and $(0.26,0)$ (red curves),
produced by the application of the kick $\mathbf{k}\equiv
(k_{x},k_{y})=(+0.25,0)$ (a1) and $(+0.5,0)$ (a2), respectively. (b1,b2)
Trajectories of the AVSs with $\left( X_{\mathrm{ps}}^{\mathrm{out}},Y_{%
\mathrm{ps}}^{\mathrm{out}}\right) =(0.12,0)$ (b1) and $(0.26,0)$ (b2),
kicked by $\mathbf{k}\equiv (k_{x},k_{y})=(+0.25,0)$ (blue curves), $%
(-0.25,0)$ (red curves) and $(0,+0.25)$ (green curves). (c1,c2) Trajectories
of the AVSs with $\left( X_{\mathrm{ps}}^{\mathrm{out}},Y_{\mathrm{ps}}^{%
\mathrm{out}}\right) =(0.12,0)$ (c1) and $(0.26,0)$ (c2) kicked by $\mathbf{k%
}\equiv (k_{x},k_{y})=(+0.5,0)$ (blue curves), $(-0.5,0)$ (red curves) and $%
(0,+0.5)$ (green curves). }
\label{TraAVS}
\end{figure}

To the best of our knowledge, such a species of AVS was not previously
reported in other\ 2D nonlinear models. Thus, we conclude that, for given
total norm $N$, instead of the single stable SVS (alias semi-vortex), which
was recently found in the system combining the SOC and contact attractive
interactions \cite{Sakaguchi}, there exists a \emph{continuous family} of
AVSs, parameterized by the offsets of the vortex' pivot, $X_{\mathrm{ps}}^{%
\mathrm{out}}$ and $Y_{\mathrm{ps}}^{\mathrm{out}}$, that take values from $0
$ to $X_{\mathrm{ps}}^{\mathrm{max}}$. The calculation of the total energy (%
\ref{Energy}) [see Fig. \ref{ASVSs}(b1)] demonstrates that its values are
almost constant for all the AVSs with the same norm\textbf{, }which seems as
continuous degeneracy of the ground state (very small fluctuations in this
figure, at a level of $<10^{-5}$, are produced by numerical-truncation
errors, which may be safely disregarded safely). Such continuous degeneracy
suggests a possibility of the existence of a hidden symmetry, which,
however, we were not able to find. A rough explanation to the degeneracy is
that the nonlocal character of the DDI may smoothen the difference in the
energy for the modes with various spatial structures. These results suggest
more possibilities for the creation of stable 2D solitons in the experiment.

\subsection{Mobility of the asymmetric vortex solitons}

Similar to what was done above for SVSs, the mobility of the AVSs was
numerically studied by applying kick to it, as per Eq. (\ref{kick}). The numerical simulations demonstrate that the computed momentums, which are defined as per Eq. (\ref{Mom}), for the moving AVSs remain constant too. Typical
examples of the respective mobility are displayed in Fig. \ref{TraAVS} by
means of trajectories of the c.o.m of the AVS's fundamental component, which
are defined as per Eq. (\ref{com2}).

Figure \ref{TraAVS} demonstrates that the
dynamics of the kicked AVSs is more complex than that of their SVS
counterparts. First, Fig. \ref{TraAVS}(a1) shows that the application of the
horizontal kick to the soliton with $X_{\mathrm{ps}}^{\mathrm{out}}\neq 0$
induces not only horizontal motion, but also motion in the vertical
direction, whose velocity increases with the increase of $X_{\mathrm{ps}}^{%
\mathrm{out}}$. Next, the shape of the trajectories initiated by a strong
kick becomes completely different. For example, in panel Fig. \ref{TraAVS}%
(a2), the trajectory of the SVSs with $X_{\mathrm{ps}}^{\mathrm{out}}=0$ has
a \textquotedblleft lacy" form (cf. blue color curve), directed in the negative $x$-direction,
while the AVS with $X_{\mathrm{ps}}^{\mathrm{out}}=0.12$ moves around the
origin along a nearly closed figure-of-eight trajectory (cf. green color curve), and the the
trajectory of the AVS with $X_{\mathrm{ps}}^{\mathrm{out}}=0.26$  changes
into a \textquotedblleft zipper" directed along the diagonal formed by
positive $x$ and negative $y$ directions (cf. red color curve). Further, the motion of
the kicked AVSs shows stronger anisotropy than the SVSs when the kick is
applied along the horizontal and vertical directions. Typical examples of
this are displayed in Figs. \ref{TraAVS}(b1, b2, c1, c2) [panels (b1,c1)
correspond to $X_{\mathrm{ps}}^{\mathrm{out}}=0.12$ and panels (b2, c2) to $%
X_{\mathrm{ps}}^{\mathrm{out}}=0.26$]. The green trajectories in
these panels pertain to the kick applied along the positive $y$-direction,
while the blue and red trajectories pertain to the kicks along the positive
and negative $x$-directions, respectively. In the latter case (the
horizontal kicks), the blue and red trajectories feature similar shapes.
However, the vertical kick, unlike the horizontal ones, does not induce
diagonal motion; instead, the trajectories form vertical spirals.

\section{Rotation of the semi-vortex solitons}

\begin{figure}[tbp]
\centering{\label{fig7a} \includegraphics[scale=0.2]{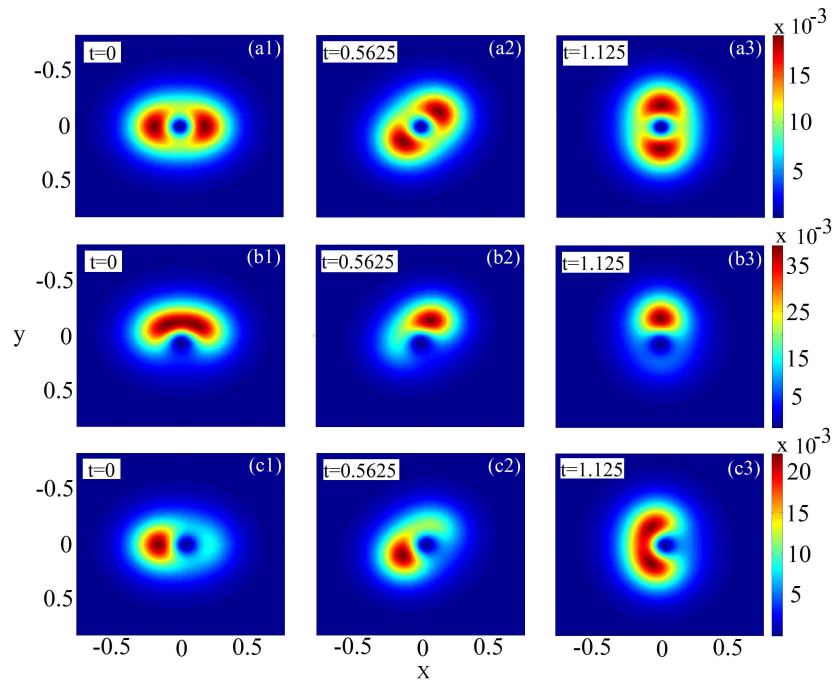}}
\caption{(Color online) The stable rotation of the vortical component of the
semi-vortex solitons driven by the rotating magnetic field. (a1-a3) The
symmetric semi-vortex. (b,c) Asymmetric semi-vortices,with $(X_{\mathrm{ps}%
}^{\mathrm{out}},Y_{\mathrm{ps}}^{\mathrm{out}})=(0.04,0)$ and $(X_{\mathrm{%
ps}}^{\mathrm{out}},Y_{\mathrm{ps}}^{\mathrm{out}})=(0,0.065)$,
respectively. The rotation speed is $\protect\omega =0.45\protect\pi $, and
the total norm of each semi-vortex soliton is $N=0.15$. The respective
evolution of the fundamental component is not shown, as it is less
conspicuous.}
\label{RotSVC}
\end{figure}

Systems in which 2D solitons are supported by the DDI make it possible to
consider dynamical generalizations of such solitons, assuming that the
external magnetic field, which polarizes the magnetic moments, rotates at a
finite angular velocity, $\omega $\textbf{,} hence the atomic moments tend
to follow it \cite{Tikhonenkov2,op28}. In the present system, the rotation
may be naturally introduced by replacing $\Theta \rightarrow \Theta +\omega
t $ in Eq. (\ref{DDI}). Figure \ref{RotSVC} shows that the magnetic
polarizations of both SVSs and AVSs can stably follow the rotation of the
driving field.

For a fixed value of the total norm, there is a critical angular speed, $%
\omega _{\mathrm{cr}}$, beyond which the soliton is not able to follow
rotation of the polarizing field, and gets destroyed (a similar effect was
reported in Ref. \cite{Tikhonenkov2,op28}). Figure \ref{ProSVS}(d2) displays
$\omega _{\mathrm{cr}}$ as a function of $N$, suggesting that the
semi-vortex solitons are more robust, following the spinning external field,
when the nonlocal nonlinearity is stronger. Indeed, being dragged by the
rotating field, the 2D anisotropic soliton must overcome the resistance of
the SOC-induced pinning, which is easier to do when the linear SOC terms are
relatively weak in comparison with the DDI.

\section{Conclusion}

The objective of this work is to construct stable 2D anisotropic solitary
vortices \ in the model of the spinor BEC composed of atoms carrying
permanent magnetic moments, with the SOC (spin-orbit coupling) applied to
the two components of the spinor wave function. The nonlinearity is
represented by the DDI (dipole-dipole interaction) between the magnetic
moments polarized in the system's plane, while the contact nonlinearity is
assumed to be negligible. Following the recent prediction of stable 2D
solitons in the form of semi-vortices in the SOC system combined with the
contact attractive interactions \cite{Sakaguchi}, we have demonstrated that
the DDI supports stable semi-vortex solitons in the present setting. The
most essential novelty in comparison with the case of the local isotropic
interactions is that, for a given total norm of the soliton, the system with
the DDI gives rise to continuous families of stable AVSs (asymmetric
(semi-)vortex solitons) parameterized by the offset of the vortex' pivot
from the soliton's fundamental component, instead of the single SVS
(symmetric (semi-)vortex soliton) found in the system with the contact
attraction. To the best of our knowledge, such composite solitons, with
linearly coupled but spatially separated fundamental and vortex components,
were not previously discovered in any other 2D nonlinear system. In
addition, the mobility of the solitons was studied by applying the kick to
them. The solitons have a negative effective mass, moving in the direction
opposite to the kick, provided that its strength exceeds the critical
(depinning) value. The critical angular velocity was found too, up to which
the 2D semi-vortices may follow the rotation of the external magnetic field
that determines the polarization of the atomic magnetic moments.

The present analysis can be extended in various directions.\ First, Ref.
\cite{Sakaguchi} suggests to look for another class of composite solitons,
namely, mixed modes, which include fundamental and vortical terms in both
components, and feature small separation between maxima of the two
components. Further, it may be interesting to consider effects of the
competition of the DDI and contact interactions, that may be added to the
system. Further, the existence of stably moving solitons suggest a
possibility to simulate collisions between them \cite{Sakaguchi}. Finally, a
challenging option is to seek for stable solitons in the 3D setting with
SOC, as suggested by the recent discovery of such 3D objects in the model
with the contact attractive nonlinearity \cite{HP}.

\begin{acknowledgments}
This work was supported, in a part, by the National Natural Science
Foundation of China through Grants 11575063 and 11547212.
\end{acknowledgments}

%

\bibliographystyle{plain}
\bibliography{apssamp}

\end{document}